\documentclass[runningheads]{llncs}

\usepackage[T1]{fontenc}
\usepackage{graphicx,verbatim}
\usepackage{hyperref}
\usepackage{amsmath}
\usepackage{amssymb}
\usepackage{xcolor}
\usepackage{multirow}
\usepackage{tabularx}
\usepackage{booktabs}

\begin{document}

\title{CT-Guided Spatially-varying Regularization for Voxel-Wise Deformable Whole-Body PET Registration}

\author{
Xiangcen Wu\inst{1} \and
Ruohua Chen\inst{2} \and
Sichun Li\inst{2} \and
Qianye Yang\inst{3} \and
Sheng Liu\inst{1} \and
Jianjun Liu\inst{2} \and
Zhaoheng Xie\inst{1}
}

\authorrunning{X. Wu et al.}

\institute{
Institute of
Medical Technology, Peking University Health Science Center, Beijing, China \\
\email{\{wuxiangcen,ls2954210845\}@bjmu.edu.cn, xiezhaoheng@pku.edu.cn}
\and
Ren Ji Hospital, Shanghai Jiao Tong University, Shanghai, China \\
\email{\{crh19870405,maozz0822,nuclearj\}@163.com}
\and
University of Oxford, Oxford, United Kingdom \\
\email{qianye.yang@eng.ox.ac.uk}
}

\maketitle
\begin{abstract}
Whole-body Positron Emission Tomography (PET) registration is essential for multi-parametric tumor characterization and assessment of metastatic disease progression. In deep learning-based deformable registration, the dense displacement field (DDF) regularizer is crucial for stabilizing optimization and preventing unrealistic deformations in large 3D volumes. A key challenge in whole-body deformable registration is anatomical heterogeneity, rigid structures (e.g., bones) should undergo stronger regularization, whereas soft tissues require more flexible deformation and weaker constraints. In this work, we propose a simple yet effective CT-guided spatially-varying regularization strategy for whole-body cross-tracer deformable PET registration. The key idea is to use the paired CT volume from the PET/CT acquisition to construct a voxel-wise regularization map for the DDF, replacing the conventional single global regularization weight. This yields anatomy-adaptive regularization strength across rigid and soft tissues. The proposed method is evaluated on a real clinical cross-tracer PET/CT dataset of 296 patients involving $^{18}$F-PSMA and $^{18}$F-FDG, showing that the proposed method achieves statistically significant improvements over weakly-supervised registration baseline in both whole-body registration performance and organ-wise alignment. Code is released at Code is released at \url{https://github.com/XiangcenWu/psma_gen}

% \href{https://anonymous.4open.science/r/WholeBodyPETRegistration-3267/}{here}.

\keywords{Image Registration  \and PET/CT \and Weakly Supervised}

\end{abstract}
% \section{Introduction}
% 1. p1, introduction to PET (claim whole-body image) registration, applications
% 2. p2, why PET (whole-body image) registration is challenging, say it comes from 2 folds: 1. misaligned semantic meaning with different contrast, hard to directly optimized by traditional intensity similarity metrics ; 2. whole body image that includes varies of organ and hard to optimize under uniform regularization. 3. Introduce related work, say how people solve this 2 challenges, in different applications, and what's their limitation, and what is under consturction in the literature.
% 3. Say in this work, we register PET using a surrogate task (CT registraition and mask registration in a weakly manner), meanwhile you proposed a regularization approach that could effectively make the regularizer adapt to varies of organs. Describe what is your hypothesis. Refer you pipeline image (\label{fig:overall_method}).
% 4. EXPLICITLY summarize your contribution 1) you proposed a whole body PET registration, 2) proposed a simple but effective CT regularizer that could adapt ddf regularizer's strength on varies of organs 3) validated the methodology on real clinical datasets with 296 patients and demonstrate the effectiveness with statistical significance; 4) and to your knowledge, is the first study that ......
% abstract intro 和 result的总结，intro4段=四段话
% last tell reader that results are sigificant
\section{Introduction}
Positron emission tomography (PET), typically acquired as hybrid PET/CT in oncologic practice, enables whole-body functional assessment of disease and plays an important role in cancer detection, staging, and therapy response monitoring~\cite{almuhaideb201118f}. In many oncologic applications, accurate image registration is essential for cross-timepoint and cross-modality analysis, enabling comparison over longitudinal scans and integration of complementary information from different imaging modalities~\cite{li2012algorithm}. Cross-tracer PET can reveal complementary biological characteristics of primary tumors and metastases that are not captured by a single tracer. For example, in prostate cancer, PET/CT scans acquired with different tracers may reflect different aspects of disease phenotype and lesion heterogeneity, making joint analysis potentially valuable for characterizing primary and metastatic disease~\cite{iacovitti2025diagnostic,chen2022added}. To support cross-tracer imaging comparison, accurate whole-body PET registration is required to establish spatial correspondence between acquisitions, particularly when PET/CT scans are obtained in separate imaging sessions and therefore are not spatially aligned.

Currently, deformable registration of whole-body PET remains highly challenging. The difficulty arises from two major aspects. First, PET scans from different tracers often have different uptake patterns. This makes conventional intensity-based similarity metrics (e.g., local cross-correlation or mean squared error) unreliable for direct optimization, particularly for deformable alignment at the voxel level. Second, whole-body images include anatomically diverse structures. Applying a single uniform deformation regularization across the entire body may overconstrain flexible regions (e.g., soft tissues) while underconstrain rigid structures (e.g., bones), leading to sub-optimal DDFs. These two factors jointly make whole-body cross-tracer PET registration more difficult than organ-specific or pure single modality registration.

To mitigate the lack of voxel-wise correspondences and inconsistent intensity semantics in multi-modality registration, weakly supervised methods leverage higher-level anatomical supervision such as segmentation masks~\cite{hu2018weakly,balakrishnan2019voxelmorph}, or landmarks \cite{matkovic2024deformable,wang2023robust} to guide learning of DDFs. Other approaches use an additional imaging modality available only during training as privileged supervision to facilitate challenging cross-modality registration without requiring that extra modality at inference time~\cite{yang2022cross}. These strategies have shown promising results in specific anatomical regions, but extending them to cross-tracer whole-body PET remains nontrivial. In a related PET/CT study, ~\cite{jonsson2022image} uses CT to guide affine and deformable registration; however, it is applied to a limited pelvis-to-neck rather than true whole-body registration. 

The challenge of how to appropriately regularize the DDF across anatomically diverse whole-body structures remains less well addressed. Spatially adaptive regularization has been explored in traditional optimization-based registration methods~\cite{pace2013locally,risser2013piecewise}. More recently, deep learning-based methods have been proposed to improve flexibility by learning a spatially-varying regularization \cite{chen2023spatially,shen2019region,niethammer2019metric,gerig2014spatially}. However, learning a voxel-wise regularization field is itself challenging, especially for whole-body registration, because the model must jointly optimize image alignment, and spatially-varying regularization weights, which can lead to unstable solutions without strong structural guidance.

\begin{figure}[h]
    \centering
    \includegraphics[width=1.0\textwidth]{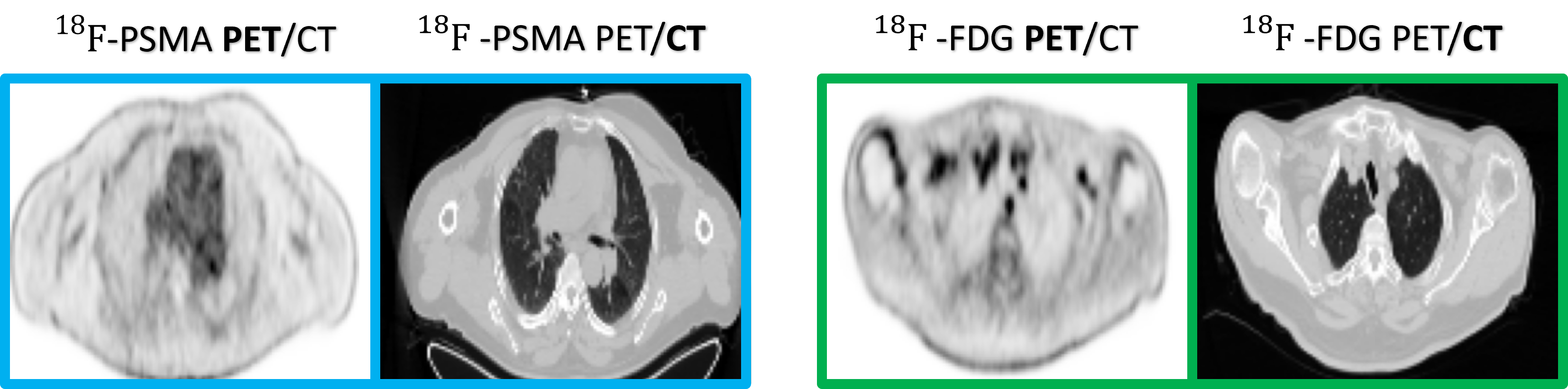} 
    \caption{Axial views from the same patient with different tracer PET/CT scans ($^{18}$F-PSMA: blue, $^{18}$F-FDG: green). PET and CT are spatially aligned within each scan session, but the two PET/CT are not aligned to each other due to scan anatomical variation.}
    \label{fig:pet-ct-align}
\end{figure}

In whole-body PET/CT, however, there is an important practical property that can be exploited. Although PET/CT examinations from different scans are not aligned to each other, the PET and CT images within each individual PET/CT scan are spatially aligned by design (Fig.~\ref{fig:pet-ct-align}). This property provides a practical advantage that we exploit in two ways: (i) CT serves as a stable anatomical surrogate for weak supervision during training, and (ii) the paired CT of the moving PET image guides spatially varying DDF regularization.

\begin{figure}[htbp] % h=here, t=top, b=bottom, p=page
    \centering
    \includegraphics[width=1.0\textwidth]{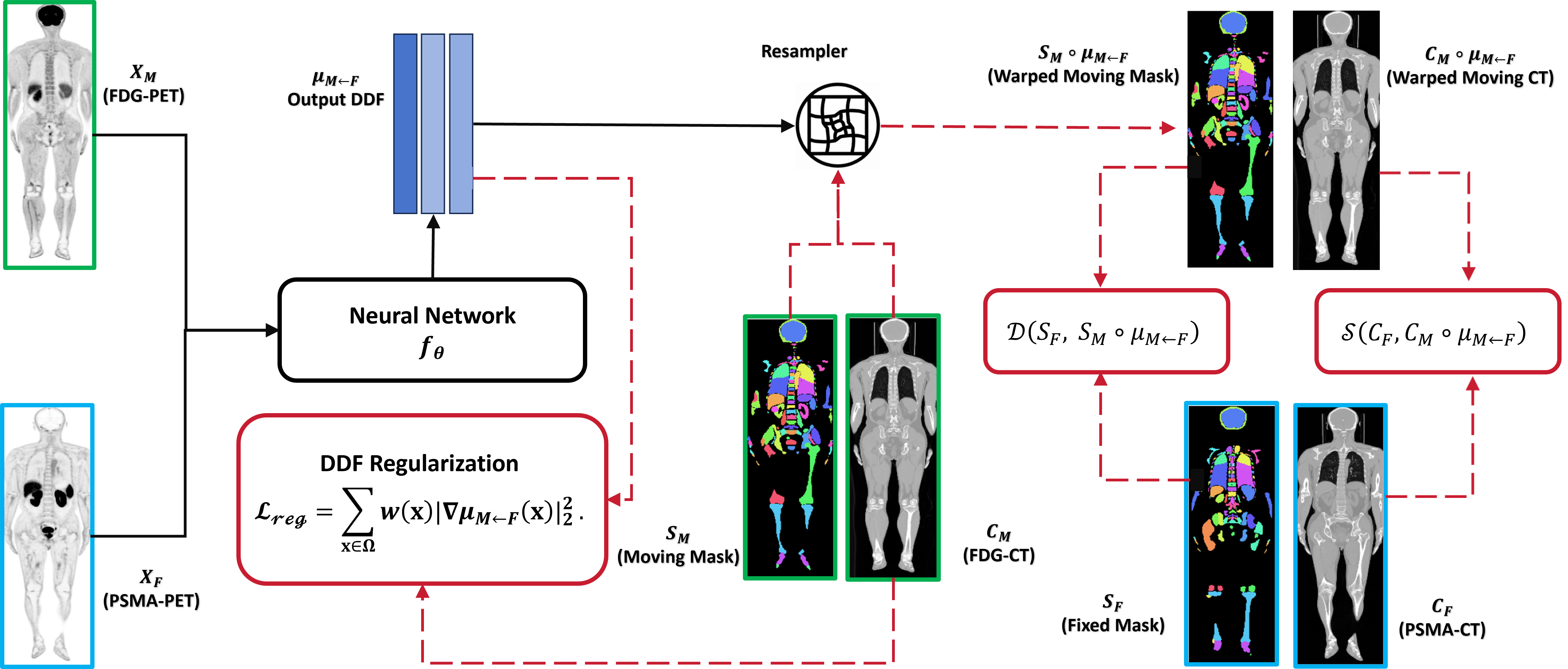} 
    \caption{Overview of the training process of the proposed framework. Blue and green boxes denote the $^{18}$F-PSMA and $^{18}$F-FDG PET/CT images and corresponding segmentation masks, respectively. Images sharing the same blue or green border are spatially aligned. Red dashed lines indicate operations used only during training. During testing, only PET images are used as input.}
    \label{fig:overall_method}
\end{figure}

Specifically, we propose a CT-guided spatially-varying regularization for deformable registration, in which the DDF regularization strength is modulated voxel-wise using anatomical information from the paired CT rather than a single global weight. In particular, higher CT Hounsfield Unit (HU) values generally correspond to denser, more rigid structures (e.g., bones), while lower HU values correspond to softer tissues~\cite{denotter2019hounsfield}; accordingly, we impose stronger penalties on the DDF regularization term (e.g., penalties on deformation field derivatives) in high-HU regions, while using weaker penalties in low-HU regions to allow greater deformation flexibility. This design enables anatomy-specific regularization strength across the whole body and accommodates heterogeneous deformation behavior across different anatomical structures. Our central hypothesis is that: (i) surrogate structural supervision from CT and weak mask correspondence can provide a stable training signal for cross-tracer whole-body PET registration, and (ii) spatially-varying regularization can improve deformation flexibility across anatomically diverse regions. An overview of the proposed pipeline is shown in Fig.~\ref{fig:overall_method}.

The main contributions of this paper are two-fold: 1) we present a whole-body deformable registration method for cross-tracer PET; 2) we propose a simple yet effective CT-guided adaptive regularization that modulates the strength of the regularization for  deformation field across different organs, improving alignment under heterogeneous anatomical motion patterns. Overall, we show that a simple, non-learned HU-value-derived voxel-wise regularization map is an effective and practical way to improve deep learning whole-body cross-tracer PET registration on a real cross-tracer cohort.

\section{Method}
For each patient, two PET/CT image sets (i.e., $^{18}$F-FDG and $^{18}$F-PSMA PET/CT studies) were acquired using different tracers. Each dataset consists of paired PET and CT volumes, and each pair is accompanied by a voxel-wise multi-anatomy segmentation mask. We define the dataset as $$\mathcal{D} = \left\{ \left(x_n^{t},\, c_n^{t},\, s_n^{t}\right) \mid n = 1, \dots, N,\; t \in \{\text{FDG}, \text{PSMA}\} \right\}$$
where $N$ denotes the total number of subjects and $t$ indicates the tracer type.

For each subject $n$ and tracer $t$, the PET and CT volumes are denoted by $x_n^{t} \in \mathbb{R}^{H \times W \times D}$ and $c_n^{t} \in \mathbb{R}^{H \times W \times D}$, respectively, while the corresponding voxel-wise segmentation map is denoted by $s_n^{t} \in \{0,1,\dots,127\}^{H \times W \times D}$. Each segmentation map contains 128 semantic classes, where each voxel is assigned a discrete label indicating its anatomical category.

\subsection{Problem Formulation}
For each subject $n$, we define the moving and fixed PET images as $X_M = x_n^{\text{FDG}}$ and $X_F = x_n^{\text{PSMA}}$, where $X_M$ and $X_F$ denote the moving (FDG-PET) and fixed (PSMA-PET) images, respectively. Each PET scan is accompanied by an aligned CT volume acquired from joint PET/CT acquisition as a tracer-independent anatomical reference, denoted by $C_M = c_n^{\text{FDG}}$ and $C_F = c_n^{\text{PSMA}}$. In addition, voxel-wise segmentation maps are available for both $X_M$ and $X_F$, denoted by $S_M = s_n^{\text{FDG}}$ and $S_F = s_n^{\text{PSMA}}$, representing the moving and fixed segmentation masks, respectively. The objective is to estimate a dense displacement field (DDF) $\mu_{M \leftarrow F} : \Omega \rightarrow \mathbb{R}^3$ that registers the FDG-PET image to the PSMA-PET image. To this end, we train a registration network $f_\theta : (X_M, X_F) \mapsto \mu_{M \leftarrow F}$, parameterized by $\theta$, which takes input of moving and fixed PET image pairs $(X_M, X_F)$ to predict a voxel-wise displacement field. The warped moving PET and CT volumes are then obtained via spatial resampling, e.g., $\tilde{X}_M = X_M \circ \mu_{M \leftarrow F}$, where $\circ$ denotes trilinear interpolation.

\subsection{Weakly Supervised Registration Loss}
Direct similarity maximization between PET volumes is challenging as $^{18}$F-FDG and $^{18}$F-PSMA exhibit distinct tracer uptake mechanisms. Instead, we exploit the anatomical consistency of CT across tracers. The registration network is trained by minimizing the loss between the warped FDG-CT and the PSMA-CT, defined as $\mathcal{L}_{\text{sim}} = -\mathcal{S}\!\left(C_F,\; C_M \circ \mu_{M \leftarrow F}\right)$, where $\mathcal{S}(\cdot,\cdot)$ denotes the Dice similarity metric. In addition to CT-based similarity, we introduce another supervision term using the available voxel-wise segmentation masks. We maximize the overlap between the warped moving mask and the fixed mask, equivalently defined as minimizing $\mathcal{L}_{\text{seg}} = -\mathcal{D}\!\left(S_F,\; S_M \circ \mu_{M \leftarrow F}\right)$, where $\mathcal{D}(\cdot,\cdot)$ denotes a segmentation similarity metric.

\subsection{CT-Guided Spatially-varying Regularization}
In deformable registration, the DDF is typically regularized by penalizing its spatial derivatives, i.e., $\mathcal{R} = \sum_{\mathbf{x} \in \Omega} \left\| \nabla \mu_{M \leftarrow F}(\mathbf{x}) \right\|_2^2$, where $\nabla$ denotes the spatial gradient operator and $\Omega$ is the image domain. In the conventional formulation, this regularization term is weighted by a global scalar, yielding $\mathcal{L}_{\text{reg}} = \lambda\,\mathcal{R}$. This imposes a spatially uniform regularization strength across the entire volume.

To incorporate anatomical priors into deformation modeling, we propose a spatially-varying regularization strategy derived from the CT volume. Let $C_M(\mathbf{x})$ denote the moving CT intensity at voxel $\mathbf{x}$. We first normalize CT intensities to the range $[0,1]$ as $$\hat{C}(\mathbf{x}) = \dfrac{C_M(\mathbf{x}) - C_M^{\min}}{C_M^{\max} - C_M^{\min}}$$
where $C_M^{\min}$ and $C_M^{\max}$ are the minimum and maximum intensities of the moving CT volume, respectively. We use the moving CT to regularize DDF as the DDF is defined on (and samples from) the moving image grid during warping. 

To allow flexible control over anatomical contrast, we then apply a nonlinear mapping with exponent $\gamma$, i.e., $\tilde{C}(\mathbf{x}) = \hat{C}(\mathbf{x})^{\gamma}$. The mapped CT is subsequently projected to a bounded interval centered at a predefined mean regularization weight $\mu_r$ with margin $\delta$, given by $w(\mathbf{x}) = (\mu_r - \delta) + 2\delta\,\tilde{C}(\mathbf{x})$, such that $w(\mathbf{x}) \in [\mu_r - \delta,\; \mu_r + \delta]$. The resulting weight map $w(\mathbf{x})$ is used as a voxel-wise regularization coefficient. The proposed regularization term is therefore defined as $$\mathcal{L}_{\text{reg}} = \sum_{\mathbf{x} \in \Omega} w(\mathbf{x}) \left\| \nabla \mu_{M \leftarrow F}(\mathbf{x}) \right\|_2^2$$

Compared with conventional global weighting, the proposed formulation enforces an adaptive DDF regularization strength, which is modulated according to $C_M$. Finally, the overall training objective is defined as $$\mathcal{L}_{\text{total}} = \mathcal{L}_{\text{sim}} + \mathcal{L}_{\text{seg}} + \mathcal{L}_{\text{reg}}$$

\begin{figure*}[h]
    \centering
    \includegraphics[width=\textwidth]{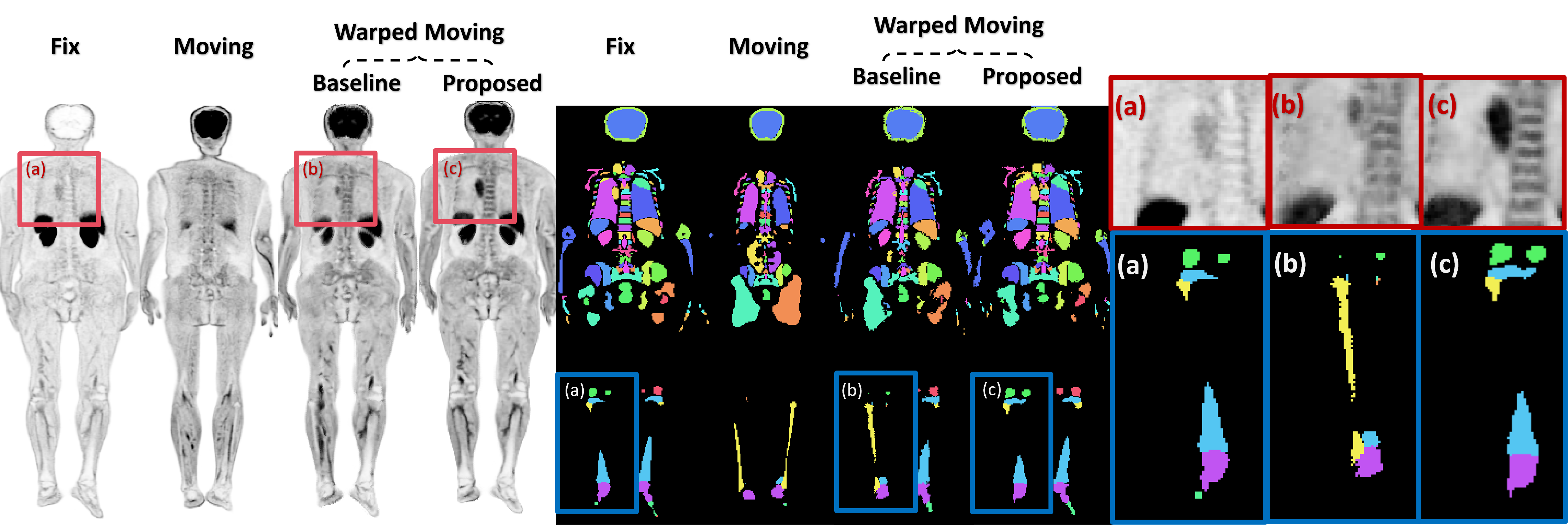}
    \caption{Left: PET images; right: corresponding segmentation masks. The fixed image is the PSMA-PET scan, and the moving image is the FDG-PET scan. Warped moving results from the baseline and proposed methods are shown for visual comparison. Insets (a)–(c) highlight representative regions of interest.}
    \label{fig:image_results}
\end{figure*}
\section{Experiments}
\subsubsection{Dataset}
We developed our method on a private dataset comprising 296 patients. Each patient underwent two separate PET/CT scans, namely $^{18}$F-FDG PET/CT and $^{18}$F-PSMA PET/CT, yielding four volumes per subject (two PET volumes and two CT volumes). Within each scan, the PET and CT volumes were spatially aligned by design due to joint PET/CT acquisition. All data were acquired using the uEXPLORER total-body PET/CT scanner (United Imaging Healthcare). Voxel-wise segmentation masks were generated on the CT modality using TotalSegmentator~\cite{wasserthal2023totalsegmentator,isensee2021nnu}, producing annotations for 128 anatomical structures for each scan. Based on the whole body segmentation mask, both CT and PET volumes were cropped to the relevant anatomical region and then resized to $128 \times 128 \times 384$ then intensity-normalized to [0,1] for model training and evaluation.

\subsubsection{Implementation Details}
We used 216 cases for training and 80 cases for testing. All experiments were conducted using the same backbone model, Swin Transformer-V2~\cite{hatamizadeh2021swin}. The model was trained for 350 epochs with a batch size of 2. In each iteration, 10 out of 128 organ masks were randomly sampled for segmentation loss computation. We used the AdamW optimizer with a learning rate of $1\times10^{-5}$. For weakly supervised learning, both the CT-based supervision term $\mathcal{L}_{\text{sim}}$ and the segmentation supervision term $\mathcal{L}_{\text{seg}}$ were implemented using the Dice loss for optimization.

\subsubsection{Evaluation Metrics}
We evaluated registration performance using three metrics: (1) Mutual Information (MI), which is reported on PET as a tracer-agnostic intensity similarity measure for cross-tracer alignment; (2) Dice score, which quantifies the degree of overlap between anatomical segmentation masks after registration; and (3) Target Registration Error (TRE, mm), computed as the Euclidean distance (in physical space) between the centroids of the registered moving segmentation and the fixed segmentation masks.

\begin{figure*}[h]
    \centering
    \includegraphics[width=\textwidth]{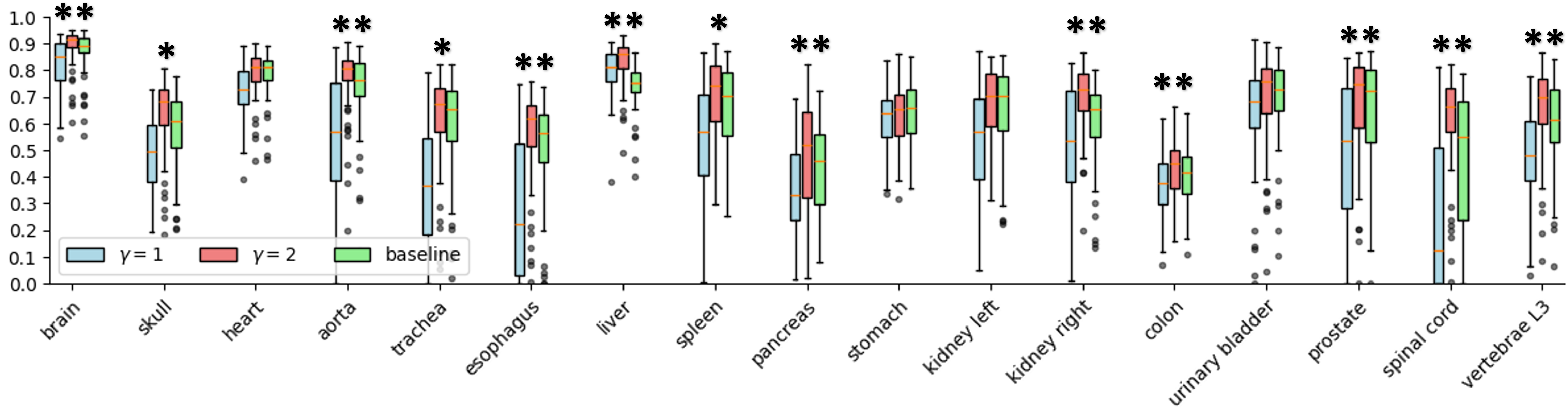}
    \caption{Boxplots of organ segmentation mask Dice scores after registration for different methods. Statistical significance markers compare $\gamma=2$ against the baseline: ** denotes $p<0.001$ and * denotes $p<0.05$.}
    \label{fig:wholebody_results}
\end{figure*}

\begin{table*}[h]
\centering
\begin{tabular}{c|cc|cc|cc}
\hline
\multirow{2}{*}{$\mu_r$}
& \multicolumn{2}{c|}{\textbf{MI$\uparrow$}}
& \multicolumn{2}{c|}{\textbf{Dice$\uparrow$}}
& \multicolumn{2}{c}{\textbf{TRE (mm)$\downarrow$}} \\
& Baseline & Proposed
& Baseline & Proposed
& Baseline & Proposed \\
\hline
$4\times 10^{3}$   & $0.60 \pm 0.11$ & $\mathbf{0.61 \pm 0.12}$
                   & $0.52 \pm 0.11$ & $0.48 \pm 0.14$
                   & $8.00 \pm 3.94$ & $11.08 \pm 7.18$ \\
$4.5\times 10^{3}$ & $0.60 \pm 0.11$ & $\mathbf{0.65 \pm 0.10}$
                   & $0.52 \pm 0.12$ & $\mathbf{0.58 \pm 0.11}$
                   & $8.37 \pm 4.81$ & $\mathbf{6.04 \pm 2.93}$ \\
$5\times 10^{3}$   & $0.54 \pm 0.13$ & $\mathbf{0.61 \pm 0.13}$
                   & $0.43 \pm 0.15$ & $\mathbf{0.47 \pm 0.16}$
                   & $13.06 \pm 8.66$ & $\mathbf{11.92 \pm 8.33}$ \\
$5.5\times 10^{3}$ & $0.56 \pm 0.13$ & $\mathbf{0.62 \pm 0.10}$
                   & $0.46 \pm 0.15$ & $\mathbf{0.53 \pm 0.10}$
                   & $11.78 \pm 7.82$ & $\mathbf{7.43 \pm 3.24}$ \\
$6\times 10^{3}$   & $0.57 \pm 0.12$ & $\mathbf{0.60 \pm 0.10}$
                   & $0.50 \pm 0.13$ & $\mathbf{0.54 \pm 0.11}$
                   & $9.86 \pm 6.08$ & $\mathbf{6.88 \pm 3.43}$ \\
$6.5\times 10^{3}$ & $0.58 \pm 0.10$ & $0.57 \pm 0.12$
                   & $0.53 \pm 0.11$ & $0.49 \pm 0.13$
                   & $6.95 \pm 3.39$ & $9.57 \pm 6.31$ \\
$7\times 10^{3}$   & $0.56 \pm 0.13$ & $0.56 \pm 0.13$
                   & $0.46 \pm 0.16$ & $0.44 \pm 0.15$
                   & $12.83 \pm 8.49$ & $12.80 \pm 8.41$ \\
\hline
\end{tabular}
\caption{Comparison of baseline and proposed methods across different $\mu_r$ values (mean $\pm$ std, $n=80$). Proposed results are shown in bold only when they are both statistically significant (paired $t$-test, $p<0.05$) and better than the baseline.}
\label{tab:mi_dice_tre_results}
\end{table*}

\subsubsection{Ablation Study Setup}
To analyze the effect of the proposed regularization terms, we conducted a controlled ablation study on top of the baseline model. The baseline uses only a global regularization coefficient $\mu_r$. The proposed variants augment the baseline with an additional term controlled by $\delta$, where $\delta$ was fixed to $3000$, and we further evaluated two values of $\gamma$: $\gamma=1$ and $\gamma=2$. For organ-wise analysis (Fig~\ref{fig:wholebody_results}), we report boxplots of Dice scores for three settings: baseline ($\mu_r$ only), $\gamma=1$ with $\delta=3000$, and $\gamma=2$ with $\delta=3000$. For the comparison in Table~\ref{tab:mi_dice_tre_results}, we compared the baseline (uniform $\mu_r$ only) and the proposed method ($\gamma=2$, $\delta=3000$). Dice and TRE are reported as
averages computed over 128 organ segmentation masks.

\section{Results and Discussion}
\subsubsection{Effect of Spatially-varying Regularization}
Table~\ref{tab:mi_dice_tre_results} summarizes the quantitative comparison between the baseline and the proposed method across different values of $\mu_r$. Overall, the proposed method improves registration performance over a broad mid-range of $\mu_r$, with the most consistent gains in MI, Dice, and TRE observed at $\mu_r=4.5\times10^3 - 6.0\times10^3$. The best overall performance is achieved at $\mu_r=4.5\times10^3$, where MI improves from $0.60\pm0.11$ to $0.65\pm0.10$, Dice from $0.52\pm0.12$ to $0.58\pm0.11$, and TRE decreases from $8.37\pm4.81$ mm to $6.04\pm2.93$ mm. Similar improvements at $\mu_r=5.5\times10^3$ and $6.0\times10^3$ further suggest that the proposed regularization is most effective under moderate global regularization strength. In contrast, the benefit is less consistent at the extremes: at $\mu_r=4.0\times10^3$, MI improves slightly but Dice and TRE degrade, while at higher $\mu_r$ values ($6.5\times10^3$ and $7.0\times10^3$), gains diminish or partially reverse, suggesting over-regularization. 
\subsubsection{Organ-wise Results}
Fig.~\ref{fig:image_results} shows representative whole-body registration examples, including PET images and corresponding anatomical segmentation masks. Fig.~\ref{fig:wholebody_results} provides organ-wise Dice distributions for the baseline, $\gamma=1$, and $\gamma=2$ (with $\delta=3000$). Overall, the proposed method improves Dice for multiple organs when $\gamma=2$, while $\gamma=1$ does not outperform the baseline, indicating that the nonlinear HU mapping ($\gamma=2$) is important for effective spatially varying regularization, as it better adapts the deformation constraint to anatomical heterogeneity across rigid structures and soft tissues.

\subsubsection{Discussion}
Our results support the hypothesis that CT can improve cross-tracer whole-body PET registration in two complementary ways: as a surrogate supervision signal during training and as spatially-varying regularization for DDF. Across the ablation study, the proposed CT-guided regularization improved performance over the baseline in a broad mid-range of global regularization strengths. This suggests that incorporating CT-derived anatomical priors helps the model achieve a better trade-off between alignment accuracy and deformation smoothness when the global regularization is tuned to an intermediate range.

\section{Conclusion}

We propose a weakly supervised whole-body cross-tracer PET deformable registration framework that leverages paired CT volumes and organ segmentation masks as surrogate anatomical supervision during training. By introducing a simple CT-guided spatially-varying regularization term, our method adaptively modulates deformation regularization across anatomically diverse regions and improves registration performance without learning a complex voxel-wise regularizer field. Experiments on a real cross-tracer PET/CT cohort of 296 patients demonstrate statistically significant improvements over baseline settings in global and organ-wise alignment, and provide evidence that anatomically informed regularization improves robustness for whole-body PET registration.

\bibliographystyle{splncs04}
\bibliography{references.bib}
\end{document}